%% file: main.tex
\documentclass{vldb}

\input{macros.tex}

\begin{document}
%

\title{Flexible Caching in Trie Joins}

%
%
%
%
%

\numberofauthors{1} 
%
\author{
%
%
\alignauthor
Oren Kalinsky\quad\quad\quad Yoav Etsion\quad\quad\quad  Benny Kimelfeld\\
       \affaddr{Technion -- Israel Institute Of Technology}\\
       \email{\{okalinsk@campus, yetsion@tce, bennyk@cs\}.technion.ac.il}
}

\maketitle
\begin{abstract}
  Traditional algorithms for multiway join computation are based on
  rewriting the order of joins and combining results of intermediate
  subqueries. Recently, several approaches have been proposed for
  algorithms that are ``worst-case optimal'' wherein all relations are
  scanned simultaneously. An example is Veldhuizen's Leapfrog Trie
  Join (LFTJ). An important advantage of LFTJ is its small memory
  footprint, due to the fact that intermediate results are full tuples
  that can be dumped immediately. However, since the algorithm does
  not store intermediate results, recurring joins must be
  reconstructed from the source relations, resulting in excessive
  memory traffic. In this paper, we address this problem by
  incorporating caches into LFTJ. We do so by adopting recent
  developments on join optimization, tying variable ordering to tree
  decomposition. While the traditional usage of tree decomposition
  computes the result for each bag in advance, our proposed approach
  incorporates caching directly into LFTJ and can dynamically adjust
  the size of the cache. Consequently, our solution balances memory
  usage and repeated computation, as confirmed by our experiments over
  SNAP datasets.
\end{abstract}




\input intro
\input preliminaries

\input algo

\input enum
\input eval

\input remarks

{
\bibliographystyle{abbrv} 
\bibliography{main}  
}

\end{document}

%% file: macros.tex
\usepackage[noend]{algorithmic}
\usepackage{algorithm}
\usepackage{cite}
\usepackage{csvsimple}
\usepackage{color}
\usepackage{enumitem}

\def\e#1{\emph{#1}}
\def\set#1{\{#1\}}

\def\tt#1{\mathcal{T}_{#1}}
\def\vars{\mathsf{vars}}

\newtheorem{thm}{Theorem}[section]

\newtheorem{lemma}[thm]{Lemma}
\newtheorem{proposition}[thm]{Proposition}

\newtheorem{theorem}[thm]{Theorem}

\newtheorem{examplethm}[thm]{Example}
\newenvironment{example}{\begin{examplethm}\em}
{\qed\end{examplethm}}

\newcommand{\commentdone}[1]{}

\newcommand{\speedup}[1]{#1$\times$}
\renewcommand{\tilde}[0]{$\sim$}

\newcommand{\hide}[1]{}

\newcommand{\algname}[1]{{\sf #1}}
\def\myrulewidth{3.20in}

\def\therule{\rule{\myrulewidth}{0.2pt}}

\newenvironment{algseries}[2]
{\centering\begin{figure}[#1]\begin{center}\def\thecaption{\caption{#2}}
\vskip-0.8em\begin{tabular}{p{\myrulewidth}}\therule\end{tabular}\vskip0.2em}
{\thecaption \end{center}\end{figure}}

\newenvironment{insidealg}[2]
{\normalsize\begin{insidecode}{#1}{#2}{Algorithm}}
{\end{insidecode}}

\newenvironment{insidesub}[2]
{\begin{insidecode}{#1}{#2}{Subroutine}}
{\end{insidecode}}

\newenvironment{insidecode}[3]
{
\begin{tabular}{p{\myrulewidth}}
\multicolumn{1}{c}{\rule{0mm}{3mm}{\bf #3} $\algname{#1}(\mbox{#2})$\vspace{-0.6em}}\\
\therule\vskip-0.8em\therule
\vspace{-1em}
\begin{algorithmic}[1]}
{\end{algorithmic}
\vskip-0.3em\therule
\end{tabular}}

\def\angs#1{\langle#1\rangle}

\def\T{\mathcal{T}}

\def\precpre{\prec_{\mathsf{pre}}}
\def\owner{\mathrm{owner}}
\def\adhesion{\mathrm{adhesion}}
\def\rt{\mathrm{root}}

\def\tot{\mathit{total}}
\def\ic{\mathit{intrmd}}
\def\cache{\mathit{cache}}

\def\mvd{\ensuremath{\rightarrow\!\!\!\!\rightarrow}}

\newcommand{\mypsfig}[2]{

\begin{figure}[t]
\centering
\input{#1.pspdftex}
\caption{\label{fig:#1}#2}
\end{figure}

}

\newcommand{\eat}[1]{}

%% file: intro.tex
\section{Introduction}
\label{sec:intro}

LeapFrog Trie Join (LFTJ)~\cite{DBLP:conf/icdt/Veldhuizen14} is a
multiway join algorithm introduced by
LogicBlox~\cite{DBLP:conf/sigmod/ArefCGKOPVW15} and implemented
within. It operates in a manner of \e{variable elimination} where
there is a linear order over the variables, and query results are
generated one by one by incrementally assigning values to each
variable in order. Trie indices over the relations guarantee that,
throughout execution, one efficiently determines whether the current
prefix of assignments cannot be matched against the database (we give
a detailed description of LFTJ in Section~\ref{sec:preliminaries}).
Veldhuizen~\cite{DBLP:conf/icdt/Veldhuizen14} has shown that LFTJ is
\e{worst-case optimal}. This yardstick of efficiency for join
algorithms has been introduced by Ngo et
al.~\cite{DBLP:conf/pods/NgoPRR12}, and it states that for every join
query, no algorithm can be asymptotically faster on the space of all
instances; in that work they presented the first algorithm that
is likewise optimal, later termed \e{NPRR}.

More traditional join optimization has been based on decomposing the
multiway join into smaller join queries and combining the intermediate
results. This approach has roots in Selinger's pairwise-join
enumeration~\cite{DBLP:conf/sigmod/SelingerACLP79}, and it includes
the application of Yannakakis's
algorithm~\cite{DBLP:conf/vldb/Yannakakis81} over a tree
decomposition of the
query~\cite{Gottlob:1999:HDT:303976.303979,DBLP:conf/wg/GottlobGMSS05}. The
advantage of LFTJ over the traditional approach is twofold. First,
LFTJ avoids the potential generation of intermediate results that may
be substantially larger than the final output size (which is a key
property in guaranteeing worst-case optimality). Second, LFTJ is very
well suited for in-memory join evaluation, since besides the trie
indices it has a close to zero memory consumption. Of course, memory
is required for buffering the tuples in the final result, but these
are never read and can be safely dumped to higher storage upon
need. In the case of an aggregate query (e.g., count the number of
tuples in the result), no such requirement arises.

But intermediate results have the advantage that their tuples can be
reused, and this is especially substantial in the presence of a
significant skew. In our experiments, we have found that LFTJ often
loses its advantage to the built-in caching of intermediate results of
the traditional approaches, and in particular, LFTJ is often required
to apply many repetition of computations.
The repeated traverals back and fourth on the trie index generate
excessive memory traffic, which has detrimental impact on the
performance of database systems~\cite{ailamaki99}.  For example, our
analysis of the memory load induced by LFTJ found that running a
single count 5-cycle query on the SNAP ca-GrQc data set generates over
$45\cdot 10^{9}$ memory accesses, whereas running the same query using
tree decomposition and Yannakakis's join generates less than $16\cdot
10^{9}$ accesses. (The implementation of both algorithms is discussed
in Section~\ref{sec:eval}.)

Nevertheless, it is not clear how LFTJ can cache results, since every
iteration involves a different partial assignment, and variables are
interdependent by the query atoms. Our goal in this work is to
accelerate LFTJ by incorporating caching in a way that (a) allows for
computation reuse, and (b) does not compromise its key advantages.  In
particular, our goal is to incorporate caching in LFTJ so that it can
utilize whatever memory it has as its disposal towards memoization.

To incorporate caching in LFTJ, we build on a recent development in
the theory of join optimization, relating worst-case optimality,
variable ordering and tree
decomposition~\cite{DBLP:journals/corr/KhamisNRR15,DBLP:journals/corr/JoglekarPR15,DBLP:conf/sigmod/TuR15}.
Specifically, given a multijoin query, we build a tree decomposition
(TD), find an order on the variables such that the order is
\e{compatibile} with the TD.  But unlike existing work, we do not
apply the join algorithm on each bag independently, but rather execute
LFTJ as originally designed. Yet, throughout the LFTJ execution we may
choose to cache partial assignments (based on some decisions that we
discuss later) or reuse cached results. The manner by which caching is
carried out, as explained in Section~\ref{sec:eval}, is based on the
fact that the variable ordering correlates with the TD.  In this way,
our caching is flexible (i.e., every cached item is optional), and it
does not violate the inherent benefits of LFTJ, while dramatically
reducing the memory load.
Concretely, running the 5-cycle count query described above on the
integrated algorithm generates only $1.4\cdot 10^{9}$ memory accesses,
which is over \speedup{30} fewer accesses than the original LFTJ
algorithm (and over \speedup{10} fewer accesses then tree
decomposition with Yannakakis's join).

But where does one get a TD from? The literature contains a plethora of
algorithms with different \e{quality} guarantees. The classical
graph-theoretic measure refers to the maximal size of a bag, and a
generalization to hypergraphs is based on the notion of a \e{hypertree
  width}. The optimal values of those (i.e., realizing the \e{tree
  width} and the \e{hypertree width}, respectively) are both NP-hard
problems~\cite{Arnborg1987,DBLP:conf/wg/GottlobGMSS05}, and efficient
algorithms exist for special cases and different approximation
guarantees~\cite{DBLP:journals/dam/BouchitteKMT04}. Other notions
include decompositions that approximate the minimal~\e{fractional
  hypertree width}~\cite{DBLP:journals/talg/Marx10}. Joglekar et
al.~\cite{DBLP:journals/corr/JoglekarPR15} determine first the
variable ordering (in order to guarantee correctness of computing an
expression comprising multiple operators), and then find a tree
decomposition that complies with this ordering, and has an
approximation guarantee against the minimum fractional hypertree
width.

In our case, a TD defines a caching scheme, and various factors are
likely to determine the effectiveness of this scheme. Importantly, our
caches correspond to the \e{adhesions} (parent-child intersections),
and the adhesion cardinalities are the dimensions of keys of our
caches; hence, small adhesions are likely to have higher hit
rates. Moreover, caches are more reusable in the presence of skewed
data. Hence, we prefer not to use any algorithm that generates a
single tree decomposition, but rather to explore a space of such
decompositions. We complement existing decomposition approaches with a
heuristic algorithm for enumerating TDs, tailored primarily towards
small adhesions. Once such a collection of TDs are generated, we
generate compatible orders. (In fact, our approach requires a property
stronger than compatibility, and we call it \e{strong compatibility}.)
Given a TD and a compatible order, we can use various techniques for
benefit estimation, such as the cost model of Chu et
al.~\cite{DBLP:conf/sigmod/ChuBS15}. A particular component of our
heuristic is an algorithm for enumerating graph separating sets with
polynomial delay, without repetitions, and by increasing size.

We experiment on three types of queries: paths, cycles and random
graphs, in various sizes.  In par with recent studies on join
algorithms, we base our experiments on data sets from the
SNAP~\cite{snapnets} and IMDB workloads.  Our experiments compare the
performance of LFTJ with and without caching, and Yannakakis's
algorithm over the TD (with each subquery computed separately, as
in~\cite{DBLP:conf/sigmod/TuR15,DBLP:conf/sigmod/PerelmanR15}), as
well as other various systems.
The results show consistent improvement compared to LFTJ (in orders of
magnitude on large queries), as well as general improvement compared
to the examined algorithms and systems.  As part of our experiments we
research several attributes of cached LFTJ, such as running on
different TDs and using a different cache sizes.

\subsection{Contributions}

Our contributions can be summarized as follows.
\begin{itemize}
\item We extend LFTJ with caching, without compromising the key
  benefits. Our caching is executed alongsize LFTJ, and its size can
  be determined dynamically according to memory availability.
\item We devise a heuristic approach to enumerating tree
  decompositions of a CQ; this approach favors small adhesions, and is
  based on enumerating graph separating sets by increasing size.
\item We present a thorough experimental study that evaluates the
  effect of caching on LFTJ and compares the results to
  state-of-the-art join algorithms.
\end{itemize}

%% file: preliminaries.tex
\section{Preliminaries}\label{sec:preliminaries}

In this section we give preliminary definitions and notation that we
use throughout the paper.

\subsection{Graphs}

We use both directed graphs and undirected graphs in this paper. Let
$g$ be a graph. For a subset $U$ of the nodes of $g$, we denote by
$g[U]$ the subgraph of $g$ \e{induced by} $U$; that is, the subgraph
of $g$ that consists of all the nodes of $U$ and all the edges between
nodes of $U$. If $V$ is the node set of $g$ and $S$ is a subset of
$V$, then $g-S$ denotes the subgraph of $g$ induced by $V\setminus S$.
A \e{separating set} of $g$ is a set $S$ of nodes such that $g-S$ is
disconnected.

\subsection{Conjunctive Queries}

In this paper we study the evaluation of a Conjunctive Query, or
\e{CQ} for short, and the problem of counting the number of tuples
resulting from this evaluation. As in recent work on worst-case
optimal
joins~\cite{DBLP:conf/icdt/Veldhuizen14,DBLP:conf/sigmod/NguyenABKNRR14,DBLP:conf/pods/NgoPRR12},
we focus here on \e{full CQs}, which are CQs without projection.
Formally, a full CQ is a sequence $\varphi_1,\dots,\varphi_m$ where
each $\varphi_i$ is a \e{subgoal} of the form $R(t_1,\dots,t_k)$ with
$R$ being a $k$-ary relation name and each $t_j$ being either a
constant or a variable. We denote by $\vars(\varphi_j)$ the set of
variables that occur in $\varphi_j$, and we denote by $\vars(q)$ the
union of the $\vars(\varphi_j)$ over all atoms $\varphi_j$ in $q$
(i.e., the set of all variables appearing in $q$).

Let $q$ be a full CQ. A \e{partial assignment} for $q$ is function
$\mu$ that maps every variable in $\vars(q)$ to either a constant
value or \e{null} (denoted $\bot$). If $\mu$ is a partial assignment
for $q$, then we denote by $q_{[\mu]}$ the full CQ that is obtained
from $q$ by replacing every variable $x$ is with $\mu(x)$, if
$\mu(x)\neq\bot$, and leaving $x$ intact if $\mu(x)=\bot$.

The \e{Gaifman graph} of a full CQ $q$ is the undirected graph that
has $\vars(q)$ as its node set and an edge between every two variables
that co-occur in a subgoal of $q$.

\subsection{Ordered Tree Decompositions}\label{sec:treedec}

Let $q=\varphi_1,\dots,\varphi_m$ be a full CQ. A \e{tree
  decomposition} (TD) of $q$ is a pair $\angs{t,\chi}$ where $t$ is a
tree and $\chi$ is a function that maps every node of $t$ to a subset
of $\vars(q)$, called a \e{bag}, such that both of the following hold.

\begin{itemize}
\item For every subgoal $\varphi_j$ there is a node $v$ of $t$
  such that $\vars(\varphi_j)\subseteq\chi(v)$. 
\item For every
  variable $x$ in $\vars(q)$, the nodes $v$ of $t$ with $x\in\chi(v)$
  induce a connected subtree of $t$.
\end{itemize}

An \e{ordered TD} of a full CQ $q$ is pair $\angs{t,\chi}$ defined
similarly to a TD, except that $t$ is a rooted and ordered tree. We
denote the root of $t$ by $\rt(t)$. Let $v$ be a node of $t$. We
denote by $t_{|v}$ the subtree of $t$ that is rooted at $v$ and
contains all of the descendants of $v$. An \e{adhesion} of $t$ is a
set of the form $\chi(v)\cap\chi(p)$, where $v$ is a node of $t$ with
a parent $p$. The \e{parent adhesion} of a non-root node $v$ (or
simply the adhesion \e{of} $v$) is the set $\chi(p)\cap\chi(v)$ where
$p$ is the parent of $v$, and is denoted by $\adhesion(v)$.

Let $q$ be a full CQ, and let $\angs{t,\chi}$ be an ordered TD of
$q$. The \e{preorder} of $t$ is the order $\prec$ over the nodes of
$t$ such that for every node $v$ with a child $c$ preceding another
child $c'$, and nodes $u$ and $u'$ in $t_{|c}$ and $t_{|c'}$,
respectively, we have $v\prec u\prec u'$. We denote the preorder of
$t$ by $\precpre$. For a variable $x$ in $\vars(q)$, the \e{owner bag}
of $x$, denoted $\owner(x)$, is the minimal node $v$ of $t$, under
$\precpre$, such that $x\in\chi(v)$.  We say that $\angs{t,\chi}$ is
\e{compatible} with an ordering $\angs{x_1,\dots,x_n}$ of $\vars(q)$
if for every $x_i$ and $x_j$, if $\owner(x_i)$ is a parent of
$\owner(x_j)$ then $i<j$~\cite{DBLP:journals/corr/JoglekarPR15}.  We
say that $\angs{t,\chi}$ is \e{strongly compatible} with
$\angs{x_1,\dots,x_n}$ if for every $x_i$ and $x_j$, if
$\owner(x_i)\precpre\owner(x_j)$ then $i<j$. Observe that strong
compatibility indeed implies compatibility (but not necessarily vice
versa).

\subsection{Trie Join}

\begin{algseries}{t}{\label{alg:lftj}Count over trie join}
\begin{insidealg}{TJCount}{$q,\angs{x_1,\dots,x_n},\T$}
\STATE $\tot:=0$
\FOR{$d=1,\dots,n$}
\STATE $\mu(x_d):=\bot$
\ENDFOR
\STATE $\algname{RJoin}(1)$
\STATE \textbf{return} $\tot$
\end{insidealg}
\begin{insidesub}{RJoin}{$d$}
\IF{$d=n+1$}
\STATE $\tot:=\tot+1$
\STATE \textbf{return}
\ENDIF
\FORALL{matching values $a$ for $x_d$ in $q_{[\mu]}$ and $\T$}
\STATE $\mu(x_d):=a$
\STATE $\algname{RJoin}(d+1)$
\ENDFOR
\STATE $\mu(x_d):=\bot$
\end{insidesub}
\end{algseries}

We now describe Veldhuizen's Leapfrog Trie Join (LFTJ)
algorithm~\cite{DBLP:conf/icdt/Veldhuizen14}. Our description is
abstract enough to apply to the \e{tributary join} of Chu et
al.~\cite{DBLP:conf/sigmod/ChuBS15}. Let $q=\varphi_1,\dots,\varphi_m$
be a full CQ. The execution of LFTJ is based on a predefined ordering
$\angs{x_1,\dots,x_n}$ of the $\vars(q)$. The correctness and
theoretical efficiency of LFTJ are guaranteed on every order of
choice, but in practice the order may have a substantial impact on the
execution cost~\cite{DBLP:conf/sigmod/ChuBS15}. Moreover, in our
instantiation of LFTJ we will use orderings with specific properties.

For each subgoal $\varphi_k$, LFTJ maintains a trie structure on the
corresponding relation $r$, where each level $i$ in the trie
corresponds to a variable $x_j$ in $\vars(\varphi_k)$ and holds values
that can be matched against $x_{j}$ so that whenever $x_j$ is in a
level above $x_{j'}$ it holds that $j<j'$. Moreover, every path from
root to leaf corresponds to a unique tuple of $r$ and vice versa.
Sibling values in the trie are stored in a sorted manner, and so, LFTJ
applies a sequence of sort-merge-joins as follows. Each trie holds an
iterator, which is initialized by pointing the root. First, all the
subgoals that contain $x_1$ advance their iterators in the first level
until a matching value $a$ is found (i.e., all iterators point to
$a$). The algorithm then proceeds recursively with the full CQ
$Q_{x_1/a}$, then proceeds to the next matching value, and so on,
until no matching values are found. A balanced-tree storage of the
sibling collections in the tries guarantees that alignment of the
iterators on matching attributes is done efficiently (in an amortized
sense), which in turn guarantees that LFTJ is \e{worst-case
  optimal}~\cite{DBLP:conf/pods/NgoPRR12}.
See~\cite{DBLP:conf/icdt/Veldhuizen14} for more details.

In this paper, it suffices to regard LFTJ abstractly as depicted in
the algorithm of Figure~\ref{alg:lftj}, and refer to it as \e{trie
  join}. The pseudocode does not compute the join, but in fact counts
the tuples in the join; the translation into \e{evaluation} is
straightforward, but we find the presentation more elegant for count.
Moreover, we will later experiment with both evaluation and counting
of joins. In the algorithm, the assignments $x_i/a$ are represented
using a global partial assignment $\mu$ that is updated by the
subroutine $\algname{RJoin}$ (Recursive Join). Note that in addition
to $\mu$, also global is the variable $\tot$ (which, in the end,
stores the resulting count).

%% file: algo.tex
\section{Caching in Trie Join} 
\label{sec:algo}
We now describe our proposed incorporation of caching in LFTJ.  For
simplicity, we will focus on the counting query and show how we extend
the algorithm $\algname{TJCount}$ of Figure~\ref{alg:lftj}.

\subsection{Intuition}

The general idea is the following.  Given the full CQ $q$, we first
construct an ordered TD $\angs{t,\chi}$ of $q$.  Let $v$ be a node of
$t$, let $\alpha$ be $\adhesion(v)$, and let $X$ be the set of all the
variables $x$ such that $\owner(x)$ is in the subtree $t_{|v|}$. Then
in the result of evaluating $q$ we have the multivalued dependency
$\alpha\mvd X$. Therefore, for every assignment $\mu$ to $\alpha$ we
can cache the assignments to $X$ (or their number in the case of
counting) and reuse them on the next time $\mu$ is encountered.

One way of obtaining the above caching is by computing the join for
every bag using the trie join, and then join the intermediate results
using an algorithm for acyclic joins such as that of
Yannakakis~\cite{DBLP:conf/vldb/Yannakakis81}, as done in
DunceCap~\cite{DBLP:conf/sigmod/PerelmanR15,DBLP:conf/sigmod/TuR15}. However,
we wish to control the memory consumption of our algorithm and avoid
computing full intermediate queries. So, the idea is to run the trie
join ordinarily, but then cache results as the algorithm runs,
conditioned on a choice of whether or not to cache using some
utilization function that estimates the value of caching. For this to
work, the ordered TD $\angs{t,\chi}$ needs to be strongly compatible
with $\angs{x_1,\dots,x_n}$, as defined in Section~\ref{sec:treedec}.

\subsection{Algorithm}

Our algorithm is depicted in Figure~\ref{alg:cachedlftj}. Again, the
algorithm is described for the counting problem. The algorithm is an
extension of the algorithm of Figure~\ref{alg:lftj} in the sense that
when no caching takes place, the two algorithms coincide.

The algorithm, named $\algname{CachedTJCount}$, takes as input a full
CQ $q$, an ordered TD $\angs{t,\chi}$ for $q$, a variable ordering
$\angs{x_1,\dots,x_n}$ for $q$, and a database $D$. The algorithm
returns the count $|q(D)|$.

The algorithm uses several global variables that are shared across
procedure calls.
\begin{itemize}
\item As in $\algname{TJCount}$, the global variable $\tot$ counts the
  joined tuples and $\mu$ stores the current partial assignment.
\item The counter $\ic(v)$, where $v$ is a node of $t$, stores an
  intermediate count of the assignments to the variables owned by the
  nodes in $t_{|v}$, given the assignment to $\adhesion(v)$ in the
  current iteration. More precisely, let $i$ be the maximal number
  such that $x_i$ is in the adhesion of $x_i$, and consider a partial
  assignment $\mu$ that is nonnull on precisely $x_1,\dots,x_i$.  Then
  in an iteration where $\mu$ is constructed, $\ic(v)$ will eventually
  hold the number of assignments $\mu'$ for the variables owned by the
  nodes in $t_{|v}$, such that some full assignment $\hat\mu$ for $q$
  is consistent with $\mu$. Observe that this number is the same for
  assignments $\mu$ that agree on the adhesion of $v$. The counter
  $\ic(v)$ has the correct value once we are done with the variables
  owned by $v$.
\item $\cache$ stores cached values $\cache[\alpha,\mu']$ for
  adhesions $\alpha$ and assignments $\mu'$ for $\alpha$. This value
  is obtained from $\ic(v)$ once the computation of $\ic(v)$ is done.
\end{itemize}

The algorithm $\algname{CachedTJCount}$ simply initializes the global
variables and call the subrouting $\algname{RCachedJoin}$, which is
the caching version of $\algname{RJoin}$. Next, we explain this
subrouting. The input takes not only the variable number $d$, but also
a factor $f$ that aggregates cached intermediate counts. 

The first part of the algorithm, lines 1--3, tests whether we are done
with the variable scan, and if so, adds $f$ to the total count.  Now
assume that $d\leq n$, and the currently iterated variable is
$x_d$. Let $v$ be $\owner(x_d)$ and $\alpha$ be $\adhesion(v)$. In
lines 6--12 we handle the case where we have just entered $v$ from a
different node of $t$. This is determined by testing whether $d>1$ and
the previous variables, $x_{d-1}$, has a different owner (in which
case $\owner(x_{d-1})\precpre v$ must hold). In that case, the
adhesion of $v$ is already assigned values in $\mu$ (since our TD is
strongly compatible with the variable ordering), and we check whether
we already have a cached result for this assignment. If so, then this
cached result is stored in $\cache[\alpha,\mu_{|\alpha}]$, where
$\mu_{|\alpha}$ is the restriction of $\mu$ to $\alpha$. Hence, if
$\cache[\alpha,\mu_{|\alpha}]$ is defined, we skip to the next
variable outside the subtree $t_{|v}$ with the factor multiplied by
$\cache[\alpha,\mu_{|\alpha}]$.  This skipping is where strong
compatibility is required, since it ensures that the nodes owned by
$t_{|v}$ constitute a consecutive interval in ${1,\dots,n}$.  We then
set $\ic(v)$ to the cached number and return.

Lines 13--18 are executed in the case where we have not entered a new
node of $t$ or we so did but did not get a cache hit.  In this case,
we continue as in $\algname{RJoin}$. In the case where $x_d$ is the
last variable owned by $v$ (i.e., $d=n$ or $v$ is not the owner of the
next variables $x_{d+1}$), we update the intermediate count by adding
the product of the intermediate results of the children of $v$. (Note
that this product is $1$ when $v$ is a leaf.)

Finally, lines~20--22 consider again the case where we have entered a
new node of $v$ from a previous node. At this point, we are about to
go back to the previous node, and so, we check whether we should cache
for $\alpha$ and $\mu_{\alpha}$. We will consider this choice later,
and for now treat it as a decision obtained from a black box. If we
indeed choose to cache, then the cached result
$\cache[\alpha,\mu_{|\alpha}]$ is set to $\ic(v)$. (This explains why
we need to maintain $\ic(v)$ to begin with.) Next, we give an example
of the execution.

{
\begin{algseries}{t}{\label{alg:cachedlftj} Cached count over trie
    join}
\begin{insidealg}{CachedTJCount}{$q,\angs{t,\chi},\angs{x_1,\dots,x_n},D$}
\STATE $\tot:=0$
\FOR{$d=1,\dots,n$}
\STATE $\mu(x_d):=\bot$
\ENDFOR
\FORALL{nodes $v$ of $t$}
\STATE $\ic(v):=0$ 
\ENDFOR
\STATE $\cache=\emptyset$
\STATE $\algname{RCachedJoin}(1,1)$
\STATE \textbf{return} $\tot$
\end{insidealg}
\begin{insidesub}{RCachedJoin}{$d,f$}
\IF{$d=n+1$}
\STATE $\tot:=\tot+f$
\STATE \textbf{return}
\ENDIF
\STATE $v:=\owner(x_d)$
\STATE $\alpha:=\adhesion(v)$
\IF{$v\neq\owner(x_{d-1})$} 
\STATE  $\ic(v):=0$
\IF{$\cache[\alpha,\mu_{|\alpha}]$ is defined} 
\STATE let $l$ be the maximum such that $\owner{x_l}$ is in $t_{|v}$
\STATE $\algname{RCachedJoin}(l+1,f\cdot \cache[\alpha,\mu_{|\alpha}])$
\STATE $\ic(v):=\cache[\alpha,\mu_{|\alpha}]$ 
\STATE \textbf{return}
\ENDIF
\ENDIF
\FORALL{matching values $a$ for $x_d$ in $q_{[\mu]}$ over $D$}
\STATE $\mu(x_d):=a$
\STATE $\algname{RCachedJoin}(d+1,f)$
\IF{$d=n$ or $v\neq\owner(x_{d+1})$}
\STATE let $c_1,\dots,c_k$ be the children of $v$ in $t$
\STATE $\ic(v):=\ic(v)+\prod_{i=1}^k\ic(c_i)$
\ENDIF
\ENDFOR
\STATE $\mu(x_d):=\bot$
\IF{$d>1$ and $v\neq\owner(x_{d-1})$}
\IF{$(\alpha,\mu_{|\alpha})$ should be cached}
\STATE $\cache[\alpha,\mu_{|\alpha}]:=\ic(v)$
\ENDIF
\ENDIF
\end{insidesub}
\end{algseries}
}

\mypsfig{example}{Example of a CQ (left) and its tree decomposition (right)}

\begin{example}\label{example:clftj}
  The graph on the left side of Figure~\ref{fig:example} denotes a
  query $q$ where every edge corresponds to an atom over a binary
  relation $R$ with the adjacent variables (that is, $R(x_1,x_2)$,
  $R(x_2,x_3)$, $R(x_2,x_4)$ and so on). The right side depicts a tree
  decomposition $\angs{t,\chi}$ of $q$. The bags are denoted by the
  ellipses and adhesions are written in the grey boxes. The tree is
  directed top down and ordered left to right. Observe that
  $\angs{t,\chi}$ is strongly compatible with $\angs{x_1,\dots,x_n}$,
  which is our order of choice in this example. Finally, our example
  database $D$ consists of four facts:
  \[R(1,1)\quad R(1,2)\quad R(2,1) \quad R(2,2)\] We now describe a
  step in $\algname{RCachedJoin}$. As said earlier, this procedure
  generalizes $\algname{RJoin}$ of Figure~\ref{alg:lftj}. We will
  illustrate the difference between the two.

  The first time the procedure reaches an index that changes the owner
  is for $d=3$, when it moves from the top bag to its child. Denote
  this child node by $v$. In an iteration with $d=3$, we have
  $\owner(x_3)$ is $v$, and the adhesion $\alpha$ is
  $\set{x_2}$. Suppose that $\mu_{|\alpha}(x_2)=1$. The algorithm then
  reaches line~8 and tests whether $\cache[\set{x_2},\mu_{|\alpha}]$
  is
  defined (line 6). On the first variable scan, this test is false,
  and so, the algorithm will go to line~13.  As in
  $\algname{RJoin}$, the algorithm finds assignments to $x_3$ and
  makes recursive calls. Since $x_3$ is not the last node owned by
  $v$, the test of line~16 is false.  Next, the algorithm reaches
  line~20. If it chooses to cache (for $\alpha$ and $\mu_{|\alpha})$,
  then $\ic(v)$ is cached as $\cache[\set{x_2},\mu_{|\alpha}]$. The
  value $\ic(v)$ should be $16$ at this point, since there are $16$
  assignments to $x_3,\dots,x_6$ (which are the variables owned by the
  nodes in the subtree rooted at $v$) that are consistent with $x_3$.
  The value $\ic(v)$ is determined in the recursive calls of line~15.

  The call with the above $\mu$ and $d=3$ later occurs again, and
  suppose that then 
  $\cache[\set{x_2},\mu_{|\alpha}]$ is defined. Then the test of
  line~8 is true, and the algorithm skips to the next index after the
  last in its subtree, namely $d=7$, with the factor $f$ multiplied by
  $16$ (which is the value in the cache). In this case, $d=7$, and so,
  $f$ is added to the total count.

  To understand how the intermediate results are calculated, we now
  consider a call with $d=4$. In this case, $x_4$ is the last
  variable with the owner $v=\owner(x_4)$. Therefore, the test of
  line~16 is true. Let $u_l$ and $u_r$ be the left and right leaves of
  $t$, respectively. For each match for $x_4$, the algorithm
  adds to $\ic(v)$ the product $\ic(u_l)\cdot\ic(u_r)$. The reader can
  verify that in our example, this produce is always $2\cdot 2=4$.
  This addition will take place on four assignments for $x_3$ and
  $x_4$, and so, $\ic(v)$ will eventually take the value $4\cdot
  4=16$. 
\end{example}

  \subsection{Correctness}

  The following theorem states the correctness of our algorithm. The
  proof has two steps. In the first step we
  prove, by induction on time, that whenever we complete with a
  node
  $v$, the number $\ic(v)$ is correct, that is, it stores the number
  of intermediate results for the subtree $t_{|v|}$ given the
  assignment for $\adhesion(v)$. In the second step we prove that
  every unit added to $\tot$ accounts for a unique tuple in $q(D)$
  and vice versa.

\begin{theorem}
  Let $q$ be a full CQ, $\angs{t,\chi}$ a tree decomposition for
  $q$, and $\angs{x_1,\dots,x_n}$ an ordering of $\vars(Q)$ such that
  $\angs{t,\chi}$ is strongly compatible with $\precpre$. The algorithm
  $\algname{CachedTJCount}(q,\angs{t,\chi},\angs{x_1,\dots,x_n},D)$
  returns $|q(D)|$.
\end{theorem}

\subsection{Discussion}

We now discuss some additional aspects of the algorithm. The decision
of line~21 of whether or not to cache may entail arbitrary arguments.
In our implementation we adopt a fairly naive approach: we cache only
if each assignment has a support (i.e., number of occurrences) larger
than a threshold. As we show in Section~\ref{sec:eval}, this already
gives us a great benefit; in future work we plan to investigate
caching policies in depth. Also, note that the algorithm allows for
arbitrary replacements or deletions from the cache.

$\algname{CachedTJCount}$ has been described for the task of counting,
which is simpler than actually evaluating $q(D)$. Nevertheless, the
counting variant of the algorithm entails all of the important
aspects, and evaluation would mainly differ in additional details. We
discuss those now. First, in evaluation $\ic(v)$ will contain
(representations of) tuple sets. We maintain $\ic(v)$ only it is
actually needed, which means that we decide to cache for either $v$ or
an ancestor of $v$. Second, instead of forwardning $f$ in the
recursive calls of lines~10 and 15, we forward a sequence of pointers
to the intermediate results. Effectively, this means that in the
result (which is currently $\tot$) will constitute a \e{factorized
  representation}~\cite{DBLP:journals/pvldb/BakibayevKOZ13,DBLP:journals/tods/OlteanuZ15}
that may be decomposed upon need (as we do in the comparisons of our
experimental evaluations). Similarly, the product of line~18 is
replaced with a factorized representation.

%% file: enum.tex
\section{Enumerating Decompositions}
\label{sec:enum}
An important factor in the effectiveness of the caches in the
algorithm $\algname{CachedTJCount}$ is their dimensionality, which is
determined by the size of the adhesions. Small adhesions imply that
are caches have a low dimension, and hence, the chance of a cache hit
(i.e., the assignment for the variables in the adhesion has occurred
in the past) is higher.\footnote{This statement applies to cases where
  the input relations have a small arity (e.g., a graph has binary
  relations), and less to the case of wide relations where the
  \e{hypertree decomposition} better captures cache
  effectiveness. Indeed, this section focuses on the former case, and
  we leave the latter to future work.}  There are, however, additional
criteria one may wish to apply in the choice of a TD towards
beneficial caching. For example, we would like to use adhesions such
that their corresponding subqueries have high \e{skews} in the data,
and then caching a small number of intermediate results can save a lot
of repeated computation. Moreover, we would like to have a TD that is
strongly compatible with an order that is estimated as good to begin
with.  Finally, we would like to get decompositions with a large
number of bags, so that we can manipulate many caches. Therefore,
instead of applying an algorithm that selects a single TD (aiming at
optimizing some specific cost function), we take the approach of
generating multiple TDs, estimating a cost on each, and selecting the
one with the best estimate. In this section, we describe a heuristic
algorithm that we use for enumerating a set of ``good'' TDs where
goodness is tailored towards small adhesions. We are not aware of any
nontrivial algorithm for enumerating TDs, except for special cases
(e.g., chordal graphs~\cite{DBLP:conf/isaac/MatsuiUU08}) that do not
apply here.

\subsection{Generic Decomposition}

We adopt a simple method for generating tree decompositions.  The
importance of this method is in the ability to plug to it an algorithm
for enumerating graph separating sets, as we explain in the next
section.  More particularly, the algorithm calls a method for solving
the \e{side-constrained graph separation} problem, or just the
\e{constrained separation} problem for short, which is defined as
follows. The input consists of an undirected graph $g$ and a set $C$
of nodes of $g$. The goal is to find a separating set $S$ of $g$ (that
is, a set $S$ of nodes such that $g-S$ is disconnected). In addition,
$S$ is required to have the property that at least one connected
component in $g-S$ is disjoint from $C$. Hence, $S$ is required to
separate $C$ from some nonempty set of nodes. We call $S$ a
\e{$C$-constrained} separating set.  We denote a call for a solver of
this problem by $\algname{ConstrainedSep}(g,C)$. In the next section
we will discuss an actual solver.  For convenience of presentation, we
assume that a solver returns the pair $\angs{S,U}$, where $S$ is a
$C$-constrained separating set and $U$ is the set of the nodes in the
connected components $g'$ of $g-C$, such that $g'$ intersects (i.e.,
has a nonempty intersection with) $C$. That is, $U$ is obtained from
$g-C$ by taking the union of the connected components $g'$ that
contain at least one element from $C$. If no such $g'$ exists, then we
define $U$ to be an arbitrary connected component of $g-C$. Observe
that $C\subseteq S\cup U$ holds.

{
\begin{algseries}{t}{\label{alg:treedec}Tree decomposition via adhesion selection}
\begin{insidealg}{GenericDecompose}{$q$}
\STATE $g:=$ the Gaifman graph of $q$
\STATE \textbf{return} \algname{RecursiveTD}$(g,\emptyset)$
\end{insidealg}
\begin{insidesub}{RecursiveTD}{$g,C$}
\STATE $\angs{S,U}\gets\algname{ConstrainedSep}(g,C)$
\IF{$S=\bot$}
\STATE \textbf{return} the singleton decomposition of $g$
\ENDIF
\STATE $\angs{t_0,\chi_0}:=\algname{RecursiveTD}(g[S\cup U],C\cup S)$
\STATE let $V_1,\dots,V_k$ be the connected comps.~of $g-(S\cup U)$
\FOR{$i=1,\dots,k$}
\STATE $\angs{t_i,\chi_i}:=\algname{RecursiveTD}(g[S\cup V_i],S)$
\ENDFOR
\STATE let $t$ be obtained from $t_0,t_1,\dots,t_k$ by connecting the
root of $t_0$ to the root of $t_i$ for all $i>1$
\STATE $\chi:=\cup_{i=0}^k\chi_i$
\STATE \textbf{return} $(t,\chi)$
\end{insidesub}
\end{algseries}
}

The algorithm, called $\algname{GenericDecompose}(q)$, is depicted in
Figure~\ref{alg:treedec}. It takes as input a full CQ $q$ and returns
an ordered TD of $q$. It first constructs the Gaifman graph $g$ of
$q$, and then calls the subroutine $\algname{RecursiveTD}(g,C)$ with
$C$ being the empty set of nodes.  The subroutine
$\algname{RecursiveTD}(g,C)$ takes as input a graph $g$ and a set $C$
of nodes of $g$, and returns an ordered TD of $g$ with the property
that the root bag contains all the nodes in $C$. So, the algorithm
first calls $\algname{ConstrainedSep}(g,C)$. Let $\angs{S,U}$ be the
result. It may be the case that the subroutine decides that no (good)
$C$-constrained separating set exists, and then the returned $S$ is
null (denoted $\bot$). In this case, the algorithm returns the
singleton TD that has only the nodes of $g$ as the single bag. This
case is handled in lines~1--3.

So now, suppose that the returned $\angs{S,U}$ is such that $S$ is a
$C$-constrained separating set. Denote by $V_1,\dots,V_k$ the
connected components of $g-(S\cup U)$. The algorithm is then applied
recursively to construct several ordered TDs:
\begin{itemize}
\item An ordered tree decomposition $\angs{t_U,\chi_U}$ of $g[S\cup
  U]$ (i.e., the induced subgraph of $S\cup U$), such that the root
  contains $C\cup S$ (line~4);
\item For $i=1,\dots,k$, an ordered tree decomposition
  $\angs{t_i,\chi_i}$ of $g[S\cup V_i]$, such that the
  root bag contains $S$ (lines~5--7).
\end{itemize}
Finally, in lines~8--10 the algorithm combines all of the tree
decompositions into a single tree decomposition (returned as the
result), by connecting the root of each $\angs{t_i,\chi_i}$ to
$\angs{t_U,\chi_U}$ as a child of the root. 

The following proposition states the correctness of the algorithm.

\begin{proposition}
  Let $q$ be a full CQ. $\algname{GenericDecompose}(q)$ returns an ordered TD of $q$.
\end{proposition}

\begin{example}\label{example:treed}
  We now describe the algorithm on the CQ $q$ depicted in
  Figure~\ref{fig:example} and described in
  Example~\ref{example:treed}.  The Gaifman graph $g$ of $q$ is the
  same as $q$, so we refer to $q$ as $g$. We first call
  \algname{RecursiveTD}$(g,\emptyset)$. So, suppose that the pair
  constructed in line~1 is $\angs{S,U}$ where $S=\set{x_2}$ and
  $U=\set{x_1}$. In line~4 we call the algorithm recursively with
  $g[\set{x_1,x_2}],\set{x_2}$. Note that $g[\set{x_1,x_2}]$ is simply
  an edge, and so it returns as the singletone decomposition.
  Moreover, here $k=1$ and $V_1=\set{x_3,\dots,x_6}$. In line~7 we
  call the algorithm with $g[S\cup V_1]$ and $S=\set{x_2}$. Note that
  $g[S\cup V_1]$ is the graph $g$ with $x_1$ removed. Let $\T$ be the
  TD on the right of Figure~\ref{fig:example}.  If the TD returned
  from the recursive call is $\T$ with the root removed, then $\T$ is
  the returned TD.  In the execution with the input $g[S\cup V_1]$ and
  $S=\set{x_2}$, returned values of $\algname{ConstrainedSep}$ can be
  $(\set{x_2,x_3,x_4},\set{x_5})$, $(\set{x_3,x_4},\set{x_2})$, and so
  on.
  \end{example}

  We note that, as defined, $\algname{RecursiveTD}$ may return a TD
  that contains redundancy in the form of a bag that is contained in
  another. In this case we can eliminate redundancy by eliminating
  the smaller bag and connecting its children to the larger set.

\subsection{Enumerating Constrained Cuts}
The algorithm $\algname{GenericDecompose}(q)$ of
Figure~\ref{alg:treedec} generates a single ordered TD. We transform
it into an enumeration algorithm by replacing line~1 with a procedure
that efficiently enumerates $C$-constrained separating sets, and then
executing the algorithm on every such a set. A key feature of the
enumeration is that it is done by \e{increasing size} of the
separating sets, and hence, if we stop the enumeration of separating
sets after $k$ sets have been generated (to bound the number the
generated TDs), \e{it is guaranteed that we have seen the $k$ smallest
  $C$-constrained separating sets}. So, we are left with the task of
enumerating the $C$-constrained separating sets by increasing size.
For that, we are using a well known technique for ranked enumeration
with polynomial delay.

Lawler-Murty's procedure~\cite{MURTY,LAWLER} reduces a general ranked
(or \e{sorted}) enumeration problem to an optimization problem with
simple constraints.  Roughly speaking, to apply the procedure to a
specific setting, one needs just to design an efficient solution to
the constrained optimization problem. Lawler-Murty's procedure is a
generalization of Yen's algorithm~\cite{YEN} for finding the $k$
shortest simple paths of a graph. Applying the algorithm gives us the
reduction described by the following lemma. In this lemma, a
``membership constraint'' means a constraint of the form ``the result
contains a node $v$'' or ``the result excludes a node $v$.''

\begin{lemma}
  Suppose that, given $g$ and $S$, a minimal $S$-constrained
  separating set under membership constraints can be found in
  polynomial time. Then the $S$-constrained separating sets can be
  enumerated by increasing size with polynomial delay.
\end{lemma}

So, to get our enumeration it suffices to devise an algorithm for
finding a minimal $S$-constrained separating set under membership
constraints. This can be done by a reduction to an ordinary
minimum-edge-cut problem. The proof is omitted, and will be given in
the full version of the paper. Consequently, we get the following
theorem.

\begin{theorem}
  Given $g$ and $S$, the $S$-constrained separating sets can be
  enumerated by increasing size with polynomial delay.
\end{theorem}

\subsection{Discussion}
In our implementation (described in the next section), we enumerate
TDs by bounding the maximal size of the adhesions (separating sets) in
the enumeration that replaces line~1 of
$\algname{GenericDecompose}$. To select the actual TD, we employ
heuristic cost functions that involve the size of the adhesion, the
number bags (higher is better), and the tree's depth (lower is
better). Moreover, we produce a variable ordering from each ordered TD
(so that the TD is strongly compatible with the ordering) and apply
the cost function of Chu et al.~\cite{DBLP:conf/sigmod/ChuBS15}. We
leave for future work the investigation of optimizing the TD
selection.

%% file: eval.tex
\section{Experimental Study}\label{sec:eval}

Our experimental study examines the performance benefits of our
caching in LFTJ, which we call here \e{CLFTJ} for short. The counting
version of CLFTJ is depicted in Figure~\ref{alg:cachedlftj}.  To
demonstrate that its performance is comparable (and often superior) to
common high-performance join algorithms, we also compare it to YTD
(Yannakakis and Tree Decomposition). The study explores both count
aggregation (denoted as \emph{count}) and query evaluation.  The
former computes $|q(D)|$ and the latter computes $q(D)$. Finally, we
explore the effect of a number of key parameters of CLFTJ.

\subsection{Implementations}

Our experiments are based on a vanilla implementation of
LFTJ~\cite{DBLP:conf/icdt/Veldhuizen14}\footnotemark.%
\footnotetext{We use the C++ STL \emph{map} as the underlying Trie data
	structure. Notably, this implementation adheres to the
	complexity requirements of the algorithm.}
The implementation\footnotemark
\footnotetext{We compiled the code using g++ 4.9.3 (with -O3 flag).}
of CLFTJ extends the vanilla LFTJ by integrating caches, as described
in Section~\ref{sec:algo} and depicted in Figure~\ref{alg:cachedlftj}.
STL's \emph{unordered\_map} is used for the caches, which support indices
that consist of up to two dimensions (attributes).
The selection of a TD is done as described in Section~\ref{sec:enum}.
We first consider caches that store \e{every} intermediate result, and
later study the impact of bounding the cache.

YTD is implemented by combining Yannakakis's acyclic join
algorithm~\cite{DBLP:conf/vldb/Yannakakis81} with TD, as described by
Gottlob et al.~\cite{Gottlob:1999:HDT:303976.303979}.  Each bag uses
$\algname{GenericJoin(abbrev.~GJ)}$, a worst-case optimal
algorithm~\cite{DBLP:NgoRR13}.  Furthermore, the complexity
requirement for the indices \textsf{seekLowerBound} is provided by a
binary search, which is enabled through the use of cascading vectors
for the Trie.  We order the attributes in a manner where the
Yannnakakis's join attributes will be higher in the Trie, similarly to
DunceCap~\cite{DBLP:conf/sigmod/PerelmanR15}.  We use the query
compiler of EmptyHeaded~\cite{DBLP:journals/corr/AbergerNOR15} (which
applies an algorithm similar to YTD) to generate the TD\footnotemark.
\footnotetext{We thank the EmptyHeaded team~\cite{DBLP:journals/corr/AbergerNOR15}
	for sharing the code and helping us with the setup.}
For queries
with only two bags we use a regular join since, in this case, the
Yannakakis reduction stage generates an unnecessary overhead.
Moreover, for count queries whose tree decompositions yields more than
two bags, we save the relevant result for the matching join attributes
(rather than storing full intermediate results).  Notably, we have
experimented with alternative YTD implementations, but they all proved
inferior to the one described above.

\subsection{Methodology}
\hide{We now describe the general methodology of our study.}

\subsubsection{Workloads} \label{sssec:workloads}

In par with other join algorithms, our evaluation is based, for the
most part, on datasets from the SNAP collection~\cite{snapnets}. The
datasets consist of wiki-Vote, p2p-Gnutella04, ca-GrQc, ego-Facebook
and ego-Twitter.  Since the distribution of values in SNAP dataset is
highly skewed, we also use IMDB to explore the effect of datasets
that are less skewed and whose data skew is not uniform across
attributes. To this end, we partition IMDB's \emph{cast\_info} table
into a \emph{male\_cast} and a \emph{female\_cast} tables, each with
attributes \emph{(person\_id} and \emph{movie\_id)}.

\subsubsection{Queries}

\def\rand{\textsf{-rand}}

We experiment using 3 types of queries:
\begin{itemize}[itemsep=1pt,topsep=0pt,parsep=0pt,partopsep=0pt] 
	\item
	\{3--7\}-path: find paths of lengths 3 to 7 for all possible nodes $a$ and $b$. For example, a valid 4-path can comprise $E(a,b), E(b,c), E(c,d)$.
	
	\item \{3--6\}-cycle: find cycles of length 3 to 6 (e.g., a
	4-cycle is $E(a,b),E(b,c),E(c,d),E(a,d)$).
	
	\item Random graphs: we generate random graphs using the
	Erd\"os-Reyni generator. The generator sets the number of
	nodes to $N$ and uses a probability $P$ to generate an edge
	between two nodes. The graph is undirected, contains no self
	edges, and has at most one edge between two nodes. We use only
	connected graphs with $N=\{5,6\}$ and $P=\{0.4, 0.6\}$. Random
	graph queries are denoted as $N\rand(P)$. For example,
	$5\rand(0.4)$ is a random graph where $N=5$ and $P=0.4$. For
	each set of parameters we generate six different graphs.
	
\end{itemize}

Note that we do not examine \e{clique} queries since they cannot be
decomposed and, therefore, CLFTJ will not offer any advantage over
LFTJ on this type of a query.

\begin{figure*}[t!]	
	\centering
	\vspace*{-2ex}  
	\includegraphics[width=1\textwidth,clip=true]{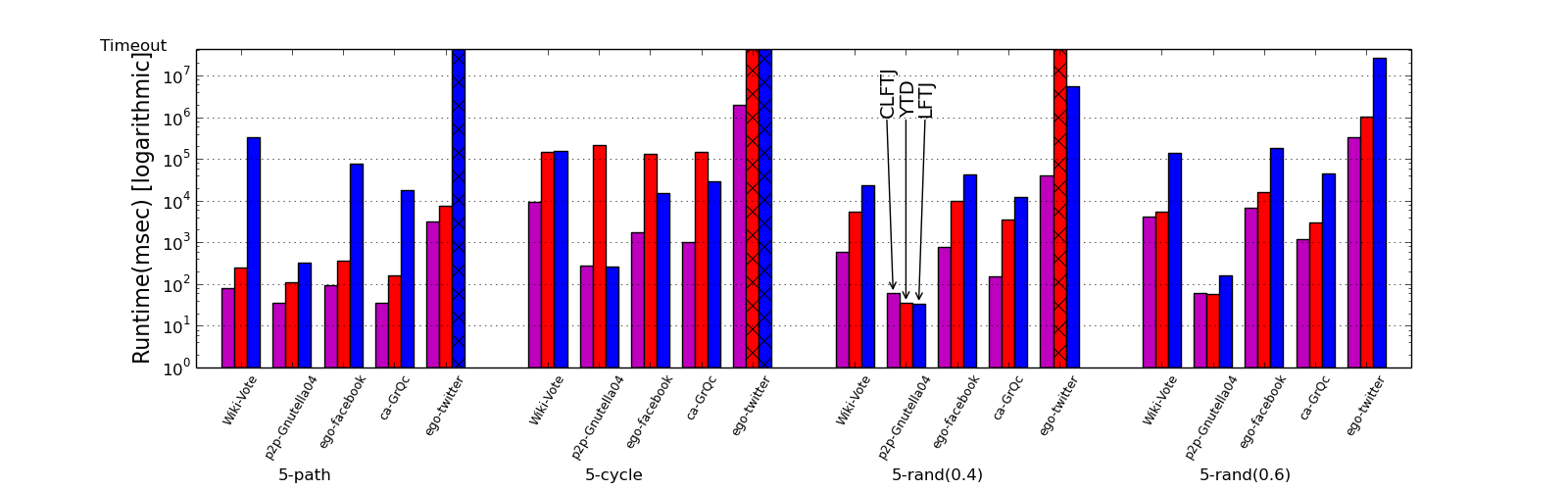}
	\caption{The runtimes observed when executing count queries using the different algorithms. YTD failed to execute 7-rand queries because its requirements exceeded the machines' memory capacity, and its results are omitted.
		Bars that represent executions that times out are marked with
		a crisscross pattern.
		\label{fig:count_compare}}
\end{figure*}

\subsubsection{Algorithms}

The main algorithms we compare against are LFTJ and YTD.  In addition
to pure algorithms, we also experiment with full systems:
\begin{itemize}[itemsep=1pt,topsep=0pt,parsep=0pt,partopsep=0pt] 
	\item \emph{System 1}(SYS1): A DBMS using a worst case-optimal join algorithm as its join engine. 
	\item \emph{System 2}(SYS2): Another DBMS using a worst case-optimal join algorithm which uses aggressive vector parallelism as its join engine.
	\item
	\emph{PostgreSQL 9.3.4} (PGSQL): An open source relational DBMS. For optimal results, the optimizer is configured to avoid merge joins and materialization.
\end{itemize}

Of course, a system has a necessarily overhead that pure algorithms do
not have. We make this comparison simply to provide a context for the
recorded running times.  We further emphasize that our experiments are
restricted to a single core, which means that we needed to
\e{restrict} the above system from utilizing our cores.

We omit other DBMSs and graph engines from our experimental study, as
they were already compared to these systems in a previous
study~\cite{DBLP:conf/sigmod/NguyenABKNRR14}.

\subsubsection{Hardware and System setup}

We use Supermicro 2028R-E1CR24N servers as our experimental
platform. Each server is configured with two Intel Xeon E5-2630 v3
processors running at 2.4 GHz, 64GB of DDR3 DRAM, and is running a
stock Ubuntu 14.0.4 Linux. 

\subsubsection{Testing Protocol}

Each experiment was run three times, and the average runtime is reported.
We set an execution timeout of 10 hours.
Executions that timed out are highlighted, and their related
speedup/slowdown is conservatively computed as if they completed the
run at the timeout mark.

\subsection{Experimental Results}

We experimented with both counting and evaluation for full CQs. We
present the results for each type separately.

\subsubsection{Count queries }
\label{sec:eval:aggs}



\begin{figure*}[t]	
	\centering
	\vspace*{-2ex}  
	\includegraphics[width=1\columnwidth,clip=true]{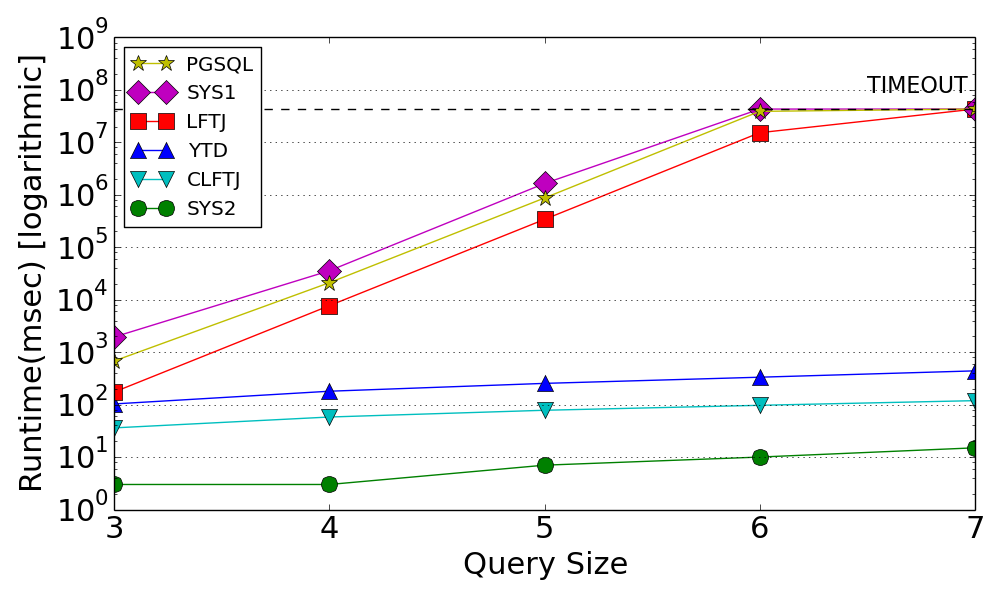}
	\includegraphics[width=1\columnwidth,clip=true]{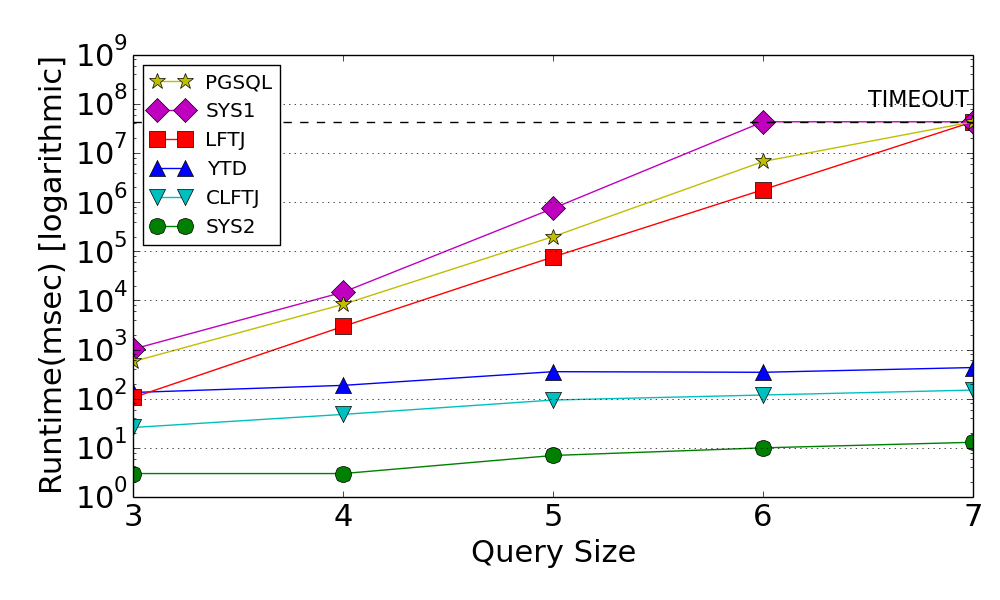}
	\vspace*{-1ex}
	\caption{The runtimes observed when executing \{3--7\}-path count queries on both the clean algorithms and DBMSs. Results shown for  wiki-Vote (left) and the ego-Facebook (right) datasets.
		\label{fig:systems_paths}}
\end{figure*}

\begin{figure*}[t]	
	\centering
	\includegraphics[width=1\columnwidth,clip=true]{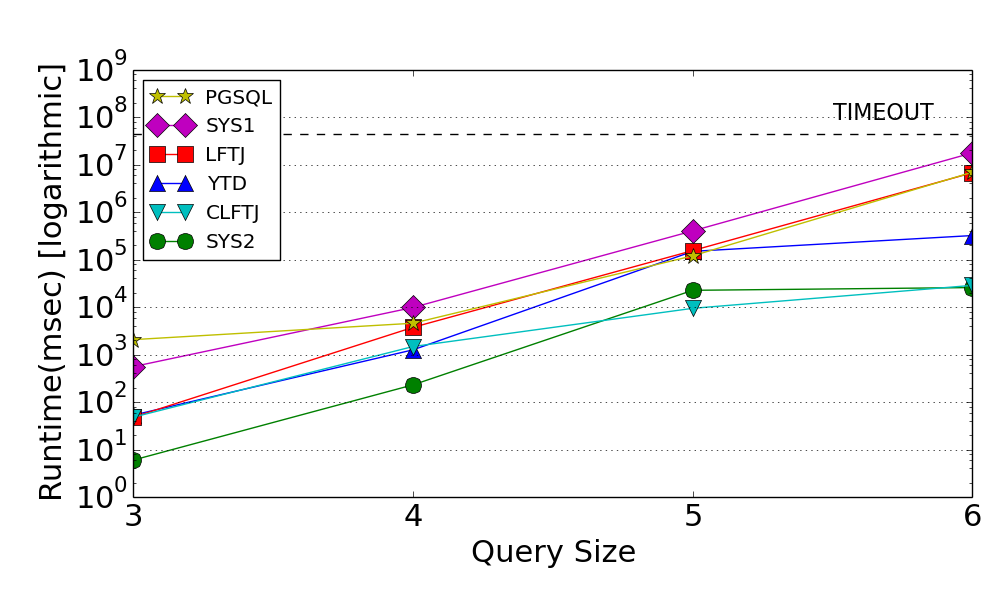}
	\includegraphics[width=1\columnwidth,clip=true]{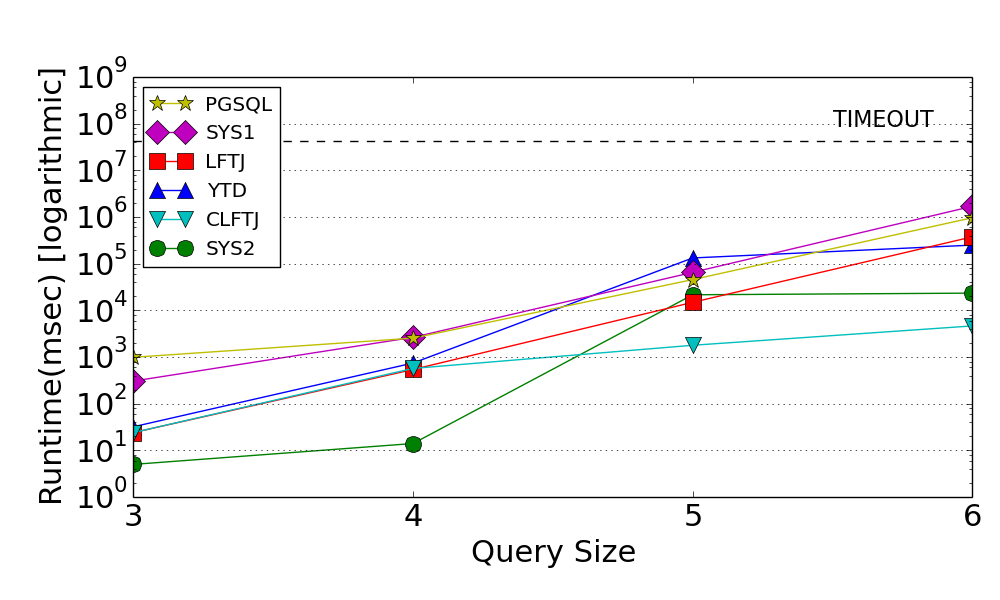}
	\vspace*{-1ex}
	\caption{
		The runtimes observed when executing \{3--6\}-cycle count queries on both the clean algorithms and DBMSs. Results shown for  wiki-Vote (left) and the ego-Facebook (right) datasets.
		\label{fig:systems_cycles}}
\end{figure*}

We first examine the performance of count queries.
Figure~\ref{fig:count_compare} presents the runtime of 5-path, 5-cycle,
and 5-rand queries on different datasets. 
The figure shows that CLFTJ is faster than the alternatives
on 5-path and 5-cycle.


The results demonstrate the effectiveness of CLFTJ when running on
datasets that are large and whose value distribution is skewed---two
properties that make them highly amenable to caching.
For example, the ego-Twitter dataset exhibits these properties and is
therefore amenable to caching. When comparing the performance of the
different algorithms, we see that CLFTJ is consistently \speedup{2--5}
faster than YTD and orders of magnitude faster than LFTJ.
On the other extreme, we have the p2p-Gnutella04 dataset that is
relatively small and whose value distribution is fairly balanced. For
this dataset, the performance benefits of CLFTJ are moderate (for
5-rand queries, both YTD and LFTJ even marginally outperform CLFTJ).


Notably, the results are consistent across different query sizes.
Figure~\ref{fig:systems_paths} presents the runtimes observed when
running \{3--7\}-path queries. (The figure also examines the
performance of full systems, which is discussed later in
Section~\ref{sec:eval:systems} below.) For brevity, we show the
results for only two of the datasets.  The figure shows that the
speedups delivered by CLFTJ over LFTJ even grows with the size of the
query. Furthermore, it shows that CLFTJ is even faster than YTD by
more than \speedup{3}.

Figure~\ref{fig:systems_cycles} examines the performance of CLFTJ for
\{3--7\}-cycle queries (again with systems discussed in
Section~\ref{sec:eval:systems}).
Again, the figure shows that CLFTJ outperforms LFTJ and YTD,
especially on larger cycle queries.  Interestingly, we see little
difference in the algorithms' running times for small, 3-cycle
queries. The reason for that is there is no tree decomposition for
triangles, and CLFTJ is effectively LFTJ.  Similarly, the performance
of CLFTJ and YTD is comparable, as YTD uses GJ for \{3--4\}-cycle
queries.

When comparing the benefits of CLFTJ over large cycle queries
(Figure~\ref{fig:systems_cycles}) and path queries
(Figure~\ref{fig:systems_paths}), we see that CLFTJ delivers better
speedups for paths.
This is attributed to the cache dimension property (the size of
adhesions). Therefore, the cache dimension for paths is set to one,
and for cycles it is set to two.  Notably, a cache whose dimension is
one shows as much more effective.  Another interesting result in the case
of 5-cycle is that YTD performs worse than LFTJ (and underperforms
CLFTJ). This is because YTD and GJ favor the opposite attributes
order, which dramatically affects its performance.  	

Finally, Figure~\ref{fig:count_compare} presents the running times for
random graph queries. 
The figure presents the results of two representative 5-rand
queries. Over all of the $5\rand(0.4)$ and $5\rand(0.6)$ queries,
CLFTJ is consistently faster by orders of magnitude than LFTJ on
average across datasets. The only exception is the p2p-Gnutella04,
which CLFTJ is slightly slower by \speedup{1.7} on two queries and
faster by \speedup{200} for the others. Compared to YTD there is a
consistent speedup of an \tilde\speedup{8}, with exception of one
query where YTD is faster by \speedup{2}. The results for 6-rand are
consistent with 5-rand, with one exception of similar runtime on
6-clique query.

\subsubsection{Query Evaluation}
\label{sec:eval:evals}

\begin{figure*}[t]	
	\centering
	\includegraphics[width=1\textwidth,clip=true]{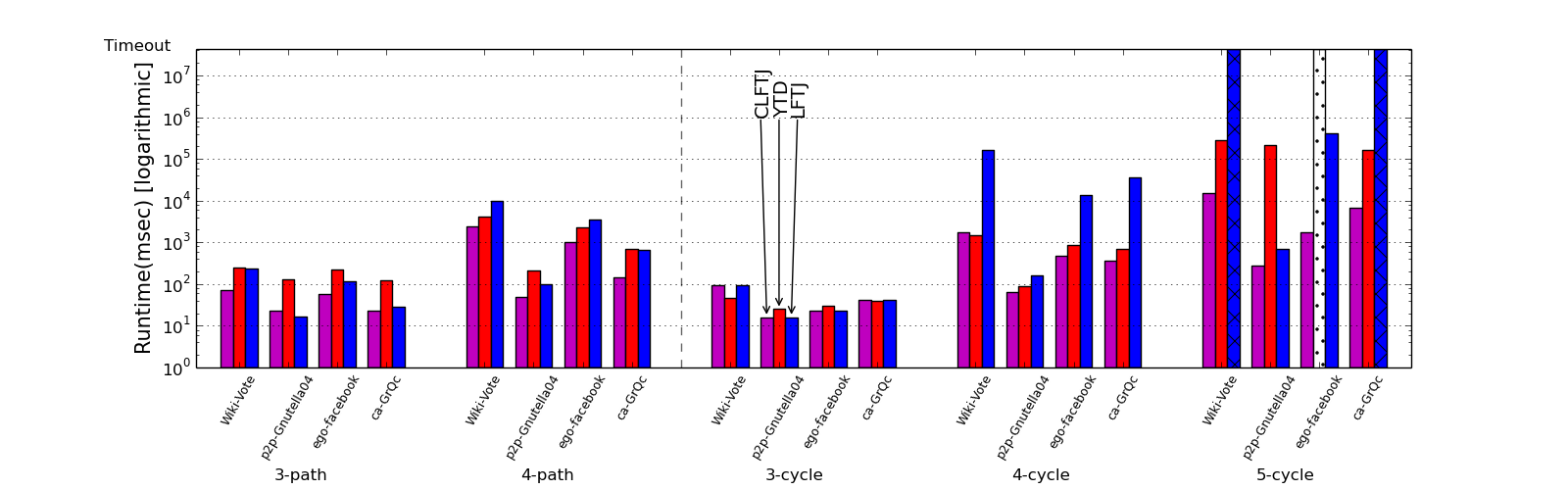}
	\caption{The runtimes of the join algorithms for full query evaluation of path and cycles queries. Queries that failed due to lack of memory are shown as white dotted bars.\label{fig:eval_path_cycle}}
\end{figure*}

\begin{figure}[t]	
	\centering
	\includegraphics[width=1\columnwidth,clip=true]{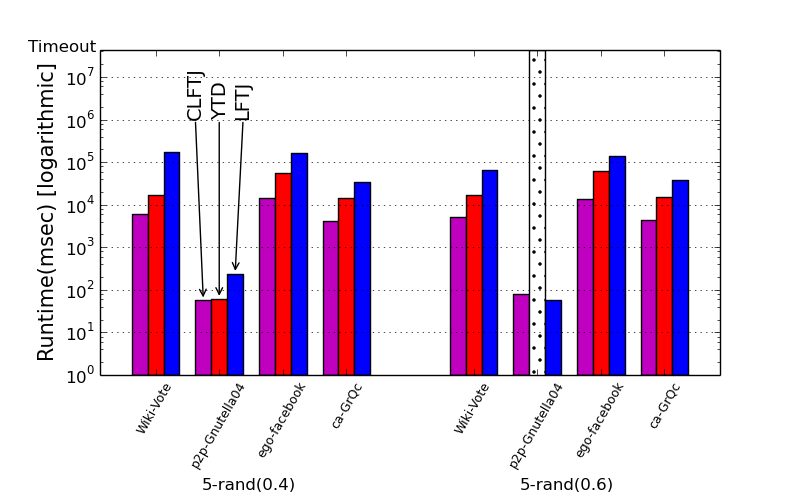}
	\caption{The runtimes of full query evaluation of random graphs. Queries that failed due to lack of memory are shown as white dotted bars.\label{fig:eval_rand}}
\end{figure}

Query evaluation produces all the tuples in the result of the query
(as opposed to counting thereof). Since our experiments measure the
total query execution runtime, including the time required to generate
the materialized result, the performance benefits of CLFTJ are
expected to be less pronounced than for count queries.
%
In contradistinction, the generation of intermediate results during query
evaluations may affect the runtime of YTD. Specifically, YTD generates
the intermediate results for all bags, even if they will not be used
in the final materialized result. In contrast, a key property of LFTJ
(and CLFTJ) is that the algorithm generates only
intermediate assignment that can be matched along with the entire
prefix assignment (according to the variable order). The performance of YTD
may thus be affected by the generation of excessive intermediate results.

Importantly, we focus our exploration of query evaluation on \e{computing} the materialized result rather then \e{storing} it, and ignore queries for which the materialized result does not fit in our machines' 64GB RAM. For this reason, we only show results for \{3--4\}-path and \{3--5\}-cycle queries, and do not discuss the ego-Twitter data set.



The results for running \{3--4\}-path query evaluations are depicted in Figure~\ref{fig:eval_path_cycle}. The figure shows
that, while gains over LFTJ are marginal for the smaller 3-path queries, CLFTJ outperforms LFTJ on the larger 4-path queries by up to \speedup{4.6} (\speedup{3.5} on average).
The performance gap is attributed to CLFTJ's caching, which captures frequently used intermediate results. In turn, this eliminates many redundant memory operations executed by LFTJ.
The CLFTJ also outperforms YTD by up to \speedup{4.6} (\speedup{3.2} on average), since the computation of YTD, which uses Yannakakis joins, becomes memory bound in the final join stages.

Figure~\ref{fig:eval_path_cycle} also presents the execution time of \{3--5\}-cycle query evaluations.
The figure shows that CLFTJ is faster than LFTJ by \speedup{2000} on average for the larger 5-cycle queries.
Interestingly, CLFTJ also proves faster than YTD by up to \speedup{800} (\speedup{280} on average) for 5-cycle queries. Similar to path queries, this performance gap is attributed to the excessive number of memory operations issued by the Yannakakis join algorithm in the final stages of the join.
%

Finally, CLFTJ also delivers performance benefits for random graphs queries.  Figure~\ref{fig:eval_rand} shows the results for representative graphs (which are consistent with the results for the other graphs).
Specifically, for $5\rand(0.4)$ queries, CLFTJ outperforms
LFTJ by \speedup{4--30}.
CLFTJ is also consistently \speedup{3--4} faster
than YTD, with the exception of p2p-Gnutella04 for which the results
are comparable.
These trends are also consistent for denser $5\rand(0.6)$ random graphs.
Here too, the results demonstrate the effectiveness of CLFTJ, whose
runtime is, on average, \tilde\speedup{10} faster than LFTJ and \tilde\speedup{4} than YTD (CLFTJ and LFTJ runtimes are comparable for p2p-Gnutella04).

\subsubsection{Dynamic Cache Size}

\begin{figure}[t]	
	\centering
	\includegraphics[width=1\columnwidth,clip=true]{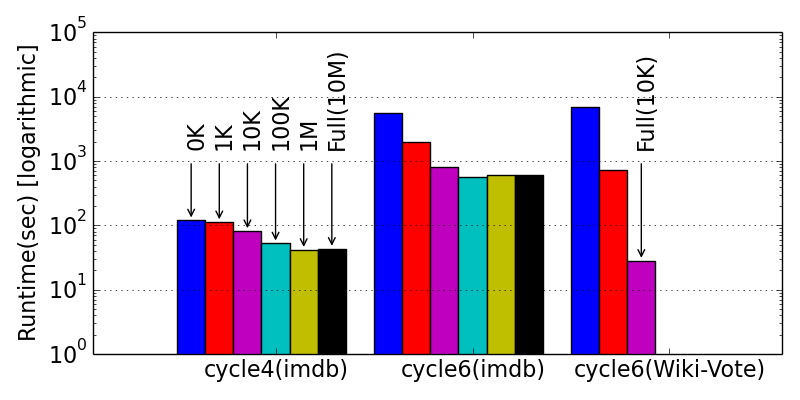}
	\caption{Different cache sizes on \{4,6\}-cycles count aggregation query over the IMDB dataset 
		and 6-cycle over wiki-Vote dataset\label{fig:cache_size}}
\end{figure}

A key benefit of LFTJ is that its memory footprint is proportional to
the original dataset and does not depend on any intermediate
results. This key property is preserved in CLFTJ through the ability
to dynamically bound its cache sizes.  Consequently, CLFTJ offers
substantial speedups when executing large queries or, alternatively,
when running in environments with limited memory resources.
Moreover, dynamic cache bounds allow CLFTJ to support multi-tenancy
of queries while preserving quality of service.

Figure~\ref{fig:cache_size} presents the runtime required to execute a
4-cycle and 6-cycle count aggregation queries (shown in
Figure~\ref{fig:imdb_cycles}) over the IMDB dataset using different
overall cache sizes.  The figure shows that the speedup provided by
CLFTJ is proportional to the overall cache sizes. Moreover, it shows
that even small caches provide substantial speedups. For example,
caching only 100K intermediate results delivers a \speedup{2.5}
speedup on 4-cycle and \speedup{7} speedup on 6-cycle, while caching
1M intermediate results provides a \speedup{3} speedup on 4-cycle and
\speedup{10} on 6-cycle.  Ultimately, caching all intermediate results
using a capacity of 10M results even incurs a small slowdown, due
to the sparse use of memory.

Figure~\ref{fig:cache_size} presents the same experiment on 6-cycle for the Wiki-Vote dataset. The Wiki-Vote dataset is much smaller and more skewed that can be fully cached with only 10K cache entries. In this case, the optimal  \speedup{246} speedup is achieved using a full cache.

We conclude that bounded caches enables CLFTJ to benefit from both
worlds. One one hand, it delivers substantial speedups over LFTJ while
preserving the bounded memory footprint property. On the other hand,
it can execute in settings where traditional join algorithms, which
store all intermediate results, either cannot execute or suffer
substantial slowdowns due to disk I/O.

\subsubsection{Tree Decomposition}

	
\begin{figure}[t]	
	\centering
	\includegraphics[width=1\columnwidth,clip=true]{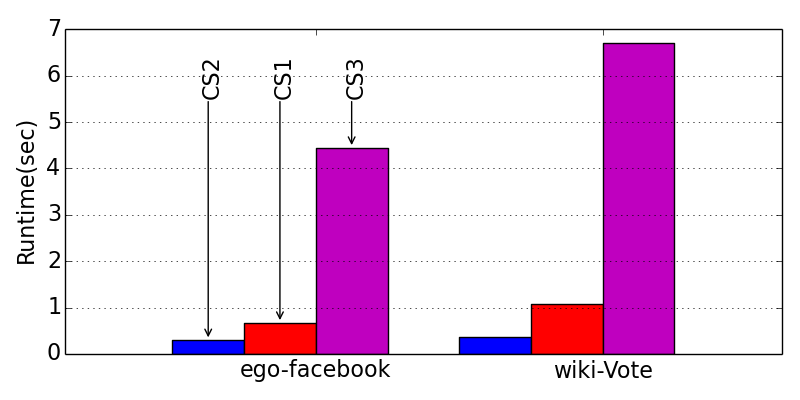}
	\caption{Runtime for the \{3,2\}-lollipop query
		(Figure~\ref{fig:lollipop}) with different cache
		structures.\label{fig:cache_dim_am}}
\end{figure}

\mypsfig{lollipop}{$\set{3,2}$-lollipop query and TDs: CS1(left), CS2(middle) and CS3(right)}
	
	
The next experiment considers the impact of orderings and
strongly-compatible TDs on the running time.  The results are in
Figure~\ref{fig:cache_dim_am}.  The figure presents the runtime of
CLFTJ on a \{3,2\}-lollipop query with different cache
structure. Importantly, due to the triangle in the lollipop graph, the
treewidth is 2. We compare the CLFTJ runtime on three cache structures
that provide the same treewidth: a single 1-dimension cache (CS1), two
1-dimension caches (CS2), and a cache structure with a single
1-dimension and a single 2-dimension caches (CS3).  The figure shows
that CS1 provides a speedup of \speedup{70--80} over LFTJ, CS2
provides a speedup of \speedup{180--190}, and CS3 only provides a
speedup of \speedup{10}.  These results demonstrate that the CLFTJ
decomposition should not target (only) small treewidth, but rather its
adhesions.

The data skew in cached attributes is another important factor that
impacts CLFTJ's performance, yet common tree decomposition algorithms
do not take data properties into account. We demonstrate the effect of
data skew on CLFTJ performance using the IMDB dataset, whose different
attributes manifest different degrees of data skew.

\begin{figure}[t]	
	\centering
	\includegraphics[width=1\columnwidth,clip=true]{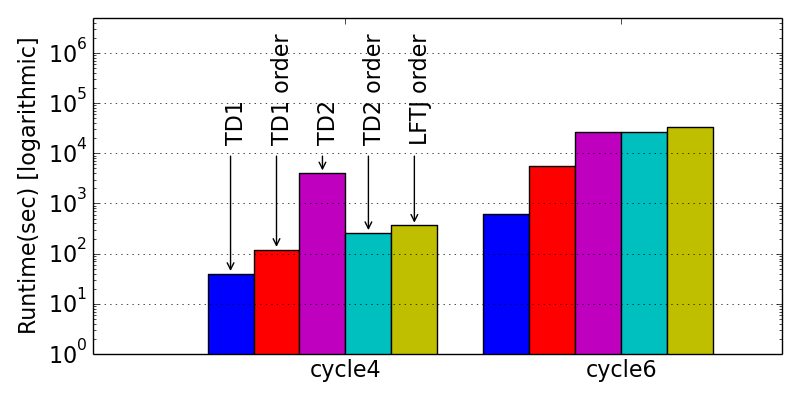}
	\caption{Comparison of CLFTJ with different TDs on 4-cycle and 6-cycle 
		and LFTJ with the imposed decompositions' attributes order 
		experimented on count aggregation on IMDB dataset 
		\label{fig:cache_skew_order} }
\end{figure}

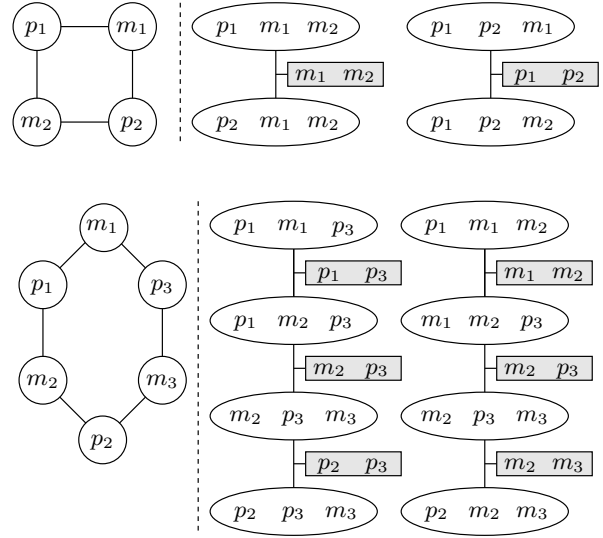
\begin{figure}[t]	
	\centering
	\input{imdb_cyc4.pspdftex}
	\vskip2em
	\input{imdb_cyc6.pspdftex}
	\caption{4-cycle (top) and 6-cycle (bottom) queries on IMDB and
		TDs: TD1 (left), TD2 (right).\label{fig:imdb_cycles}}
\end{figure}

Figure~\ref{fig:imdb_cycles} depicts two TDs, TD1 and TD2, of two
queries, 4-cycle and 6-cycle, and Figure~\ref{fig:cache_skew_order}
presents their respective runtimes. TD1 favors person\_id for caching
and TD2 favors movie\_id for caching.
While the decompositions are isomorphic (similar from a graph perspective), we see that their performance vary greatly.
The reason for the performance variation is that the person\_id attribute exhibits greater data skew than the movie\_id. It is therefore more effective to apply caches to the person\_id attribute.

Another interesting result is the performance impact of the order of
attributes. For each TD, we selected an ordering such that the TD is
strongly compatible with the ordering. Simply using LFTJ with the
imposed attribute order offers a \speedup{10} speedup over the
original LFTJ order.  Notably, a recent study by Chu et
al.~\cite{DBLP:conf/sigmod/ChuBS15} proposed a method to estimate the
cost of attributes order in LFTJ. The method estimates the cost of
TD2\_order to be \tilde\speedup{2} higher than TD1\_order. The
runtimes of the different attributes orders is shown in
Figure~\ref{fig:cache_skew_order}. Hence, in these queries the cost
function of Chu et al.~\cite{DBLP:conf/sigmod/ChuBS15} turns out to be
very beneficial as parameter of choosing the TD for caching.

\subsubsection{Comparison to Systems}
\label{sec:eval:systems}

To explore the scaling trends of the pure algorithms, compared to
those of the DBMSs, we ran the queries on PGSQL (using pair-wise join), SYS1 and SYS2 
(which are based on worst-case optimal join algorithms). For brevity, we show the
results for only two datasets: Wiki-Vote and ego-Facebook. Notably,
these are consistent with the results obtained for the other SNAP
datasets.

Figure~\ref{fig:systems_paths} shows the results for \{3--7\}-path
count queries. The first thing to note in the figure is that the
scaling of vanilla LFTJ and SYS1 are correlated.  We attribute the
\speedup{10} performance difference between the two to the overheads
associated with running a full DBMS vs.~a pure algorithm.
Importantly, the figure demonstrates that the performance benefit of
CLFTJ and YTD over LFTJ increases with the query size at an
exponential rate. Moreover, it also shows that even though CLFTJ and
YTD have similar scaling trends for path queries, CLFTJ runs almost an
order of magnitude faster.

Figure~\ref{fig:systems_cycles} depicts a similar comparison for the
queries \{3--6\}-cycle. Again, the figure shows consistent scaling
trends for the vanilla algorithms and the DBMSs that utilize them.
A comparison between SYS2 and YTD shows that SYS2 is much faster than YTD. In this case, the DBMS is
faster than a pure algorithm since its implementation is
massively parallelized using the processor's wide vector unit.  Due to
the parallel implementation, SYS2 is much faster than LFTJ on path
queries. Nevertheless, the sequential CLFTJ implementation is still
comparable to SYS2 for \{5--6\}-cycles queries (and is even faster on
some datasets).

%% file: imdb_cyc4.pspdftex
\begin{picture}(0,0)%
\includegraphics{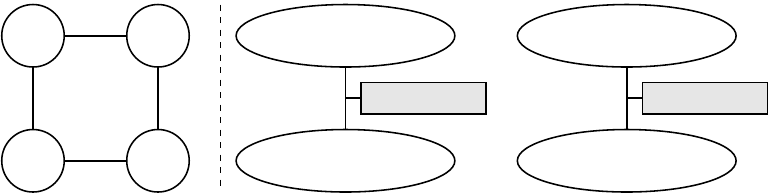}%
\end{picture}%
\setlength{\unitlength}{3947sp}%
\begingroup\makeatletter\ifx\SetFigFont\undefined%
\gdef\SetFigFont#1#2#3#4#5{%
  \reset@font\fontsize{#1}{#2pt}%
  \fontfamily{#3}\fontseries{#4}\fontshape{#5}%
  \selectfont}%
\fi\endgroup%
\begin{picture}(3695,924)(3518,-7123)
\put(3676,-6394){\makebox(0,0)[b]{\smash{{\SetFigFont{9}{10.8}{\familydefault}{\mddefault}{\updefault}{\color[rgb]{0,0,0}$p_1$}%
}}}}
\put(6751,-6694){\makebox(0,0)[b]{\smash{{\SetFigFont{9}{10.8}{\familydefault}{\mddefault}{\updefault}{\color[rgb]{0,0,0}$p_1$}%
}}}}
\put(4276,-6394){\makebox(0,0)[b]{\smash{{\SetFigFont{9}{10.8}{\familydefault}{\mddefault}{\updefault}{\color[rgb]{0,0,0}$m_1$}%
}}}}
\put(3676,-6994){\makebox(0,0)[b]{\smash{{\SetFigFont{9}{10.8}{\familydefault}{\mddefault}{\updefault}{\color[rgb]{0,0,0}$m_2$}%
}}}}
\put(4276,-6994){\makebox(0,0)[b]{\smash{{\SetFigFont{9}{10.8}{\familydefault}{\mddefault}{\updefault}{\color[rgb]{0,0,0}$p_2$}%
}}}}
\put(4876,-6394){\makebox(0,0)[b]{\smash{{\SetFigFont{9}{10.8}{\familydefault}{\mddefault}{\updefault}{\color[rgb]{0,0,0}$p_1$}%
}}}}
\put(5176,-6394){\makebox(0,0)[b]{\smash{{\SetFigFont{9}{10.8}{\familydefault}{\mddefault}{\updefault}{\color[rgb]{0,0,0}$m_1$}%
}}}}
\put(5176,-6994){\makebox(0,0)[b]{\smash{{\SetFigFont{9}{10.8}{\familydefault}{\mddefault}{\updefault}{\color[rgb]{0,0,0}$m_1$}%
}}}}
\put(4876,-6994){\makebox(0,0)[b]{\smash{{\SetFigFont{9}{10.8}{\familydefault}{\mddefault}{\updefault}{\color[rgb]{0,0,0}$p_2$}%
}}}}
\put(5476,-6394){\makebox(0,0)[b]{\smash{{\SetFigFont{9}{10.8}{\familydefault}{\mddefault}{\updefault}{\color[rgb]{0,0,0}$m_2$}%
}}}}
\put(5476,-6994){\makebox(0,0)[b]{\smash{{\SetFigFont{9}{10.8}{\familydefault}{\mddefault}{\updefault}{\color[rgb]{0,0,0}$m_2$}%
}}}}
\put(5401,-6694){\makebox(0,0)[b]{\smash{{\SetFigFont{9}{10.8}{\familydefault}{\mddefault}{\updefault}{\color[rgb]{0,0,0}$m_1$}%
}}}}
\put(5701,-6694){\makebox(0,0)[b]{\smash{{\SetFigFont{9}{10.8}{\familydefault}{\mddefault}{\updefault}{\color[rgb]{0,0,0}$m_2$}%
}}}}
\put(6226,-6394){\makebox(0,0)[b]{\smash{{\SetFigFont{9}{10.8}{\familydefault}{\mddefault}{\updefault}{\color[rgb]{0,0,0}$p_1$}%
}}}}
\put(7051,-6694){\makebox(0,0)[b]{\smash{{\SetFigFont{9}{10.8}{\familydefault}{\mddefault}{\updefault}{\color[rgb]{0,0,0}$p_2$}%
}}}}
\put(6826,-6994){\makebox(0,0)[b]{\smash{{\SetFigFont{9}{10.8}{\familydefault}{\mddefault}{\updefault}{\color[rgb]{0,0,0}$m_2$}%
}}}}
\put(6526,-6994){\makebox(0,0)[b]{\smash{{\SetFigFont{9}{10.8}{\familydefault}{\mddefault}{\updefault}{\color[rgb]{0,0,0}$p_2$}%
}}}}
\put(6526,-6394){\makebox(0,0)[b]{\smash{{\SetFigFont{9}{10.8}{\familydefault}{\mddefault}{\updefault}{\color[rgb]{0,0,0}$p_2$}%
}}}}
\put(6226,-6994){\makebox(0,0)[b]{\smash{{\SetFigFont{9}{10.8}{\familydefault}{\mddefault}{\updefault}{\color[rgb]{0,0,0}$p_1$}%
}}}}
\put(6826,-6394){\makebox(0,0)[b]{\smash{{\SetFigFont{9}{10.8}{\familydefault}{\mddefault}{\updefault}{\color[rgb]{0,0,0}$m_1$}%
}}}}
\end{picture}%

%% file: imdb_cyc6.pspdftex
\begin{picture}(0,0)%
\includegraphics{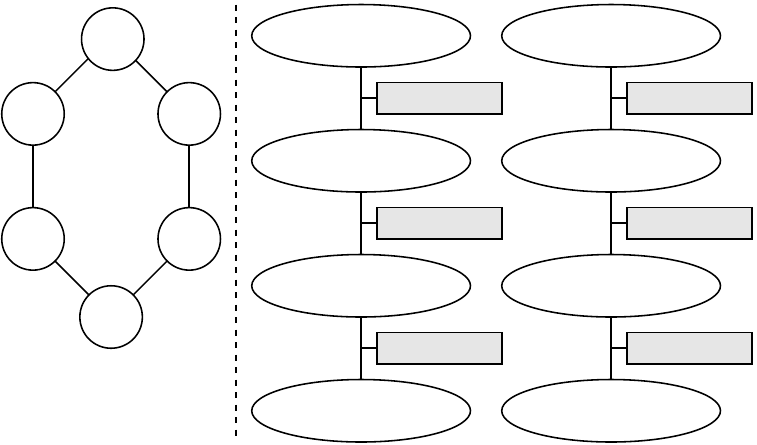}%
\end{picture}%
\setlength{\unitlength}{3947sp}%
\begingroup\makeatletter\ifx\SetFigFont\undefined%
\gdef\SetFigFont#1#2#3#4#5{%
  \reset@font\fontsize{#1}{#2pt}%
  \fontfamily{#3}\fontseries{#4}\fontshape{#5}%
  \selectfont}%
\fi\endgroup%
\begin{picture}(3620,2124)(3443,-8323)
\put(4351,-7369){\makebox(0,0)[b]{\smash{{\SetFigFont{9}{10.8}{\familydefault}{\mddefault}{\updefault}{\color[rgb]{0,0,0}$m_3$}%
}}}}
\put(6601,-7894){\makebox(0,0)[b]{\smash{{\SetFigFont{9}{10.8}{\familydefault}{\mddefault}{\updefault}{\color[rgb]{0,0,0}$m_2$}%
}}}}
\put(4351,-6769){\makebox(0,0)[b]{\smash{{\SetFigFont{9}{10.8}{\familydefault}{\mddefault}{\updefault}{\color[rgb]{0,0,0}$p_3$}%
}}}}
\put(3976,-6394){\makebox(0,0)[b]{\smash{{\SetFigFont{9}{10.8}{\familydefault}{\mddefault}{\updefault}{\color[rgb]{0,0,0}$m_1$}%
}}}}
\put(3976,-7744){\makebox(0,0)[b]{\smash{{\SetFigFont{9}{10.8}{\familydefault}{\mddefault}{\updefault}{\color[rgb]{0,0,0}$p_2$}%
}}}}
\put(3601,-6769){\makebox(0,0)[b]{\smash{{\SetFigFont{9}{10.8}{\familydefault}{\mddefault}{\updefault}{\color[rgb]{0,0,0}$p_1$}%
}}}}
\put(3601,-7369){\makebox(0,0)[b]{\smash{{\SetFigFont{9}{10.8}{\familydefault}{\mddefault}{\updefault}{\color[rgb]{0,0,0}$m_2$}%
}}}}
\put(6376,-6394){\makebox(0,0)[b]{\smash{{\SetFigFont{9}{10.8}{\familydefault}{\mddefault}{\updefault}{\color[rgb]{0,0,0}$m_1$}%
}}}}
\put(6676,-6394){\makebox(0,0)[b]{\smash{{\SetFigFont{9}{10.8}{\familydefault}{\mddefault}{\updefault}{\color[rgb]{0,0,0}$m_2$}%
}}}}
\put(5176,-7594){\makebox(0,0)[b]{\smash{{\SetFigFont{9}{10.8}{\familydefault}{\mddefault}{\updefault}{\color[rgb]{0,0,0}$p_3$}%
}}}}
\put(5401,-7294){\makebox(0,0)[b]{\smash{{\SetFigFont{9}{10.8}{\familydefault}{\mddefault}{\updefault}{\color[rgb]{0,0,0}$m_2$}%
}}}}
\put(5401,-7894){\makebox(0,0)[b]{\smash{{\SetFigFont{9}{10.8}{\familydefault}{\mddefault}{\updefault}{\color[rgb]{0,0,0}$p_2$}%
}}}}
\put(4876,-6994){\makebox(0,0)[b]{\smash{{\SetFigFont{9}{10.8}{\familydefault}{\mddefault}{\updefault}{\color[rgb]{0,0,0}$p_1$}%
}}}}
\put(5176,-6394){\makebox(0,0)[b]{\smash{{\SetFigFont{9}{10.8}{\familydefault}{\mddefault}{\updefault}{\color[rgb]{0,0,0}$m_1$}%
}}}}
\put(4876,-6394){\makebox(0,0)[b]{\smash{{\SetFigFont{9}{10.8}{\familydefault}{\mddefault}{\updefault}{\color[rgb]{0,0,0}$p_1$}%
}}}}
\put(5491,-6406){\makebox(0,0)[b]{\smash{{\SetFigFont{9}{10.8}{\familydefault}{\mddefault}{\updefault}{\color[rgb]{0,0,0}$p_3$}%
}}}}
\put(5476,-7594){\makebox(0,0)[b]{\smash{{\SetFigFont{9}{10.8}{\familydefault}{\mddefault}{\updefault}{\color[rgb]{0,0,0}$m_3$}%
}}}}
\put(4876,-7594){\makebox(0,0)[b]{\smash{{\SetFigFont{9}{10.8}{\familydefault}{\mddefault}{\updefault}{\color[rgb]{0,0,0}$m_2$}%
}}}}
\put(4876,-8194){\makebox(0,0)[b]{\smash{{\SetFigFont{9}{10.8}{\familydefault}{\mddefault}{\updefault}{\color[rgb]{0,0,0}$p_2$}%
}}}}
\put(5476,-8194){\makebox(0,0)[b]{\smash{{\SetFigFont{9}{10.8}{\familydefault}{\mddefault}{\updefault}{\color[rgb]{0,0,0}$m_3$}%
}}}}
\put(5176,-8194){\makebox(0,0)[b]{\smash{{\SetFigFont{9}{10.8}{\familydefault}{\mddefault}{\updefault}{\color[rgb]{0,0,0}$p_3$}%
}}}}
\put(6676,-7594){\makebox(0,0)[b]{\smash{{\SetFigFont{9}{10.8}{\familydefault}{\mddefault}{\updefault}{\color[rgb]{0,0,0}$m_3$}%
}}}}
\put(6601,-7294){\makebox(0,0)[b]{\smash{{\SetFigFont{9}{10.8}{\familydefault}{\mddefault}{\updefault}{\color[rgb]{0,0,0}$m_2$}%
}}}}
\put(6901,-7294){\makebox(0,0)[b]{\smash{{\SetFigFont{9}{10.8}{\familydefault}{\mddefault}{\updefault}{\color[rgb]{0,0,0}$p_3$}%
}}}}
\put(6076,-8194){\makebox(0,0)[b]{\smash{{\SetFigFont{9}{10.8}{\familydefault}{\mddefault}{\updefault}{\color[rgb]{0,0,0}$p_2$}%
}}}}
\put(6676,-8194){\makebox(0,0)[b]{\smash{{\SetFigFont{9}{10.8}{\familydefault}{\mddefault}{\updefault}{\color[rgb]{0,0,0}$m_3$}%
}}}}
\put(6076,-7594){\makebox(0,0)[b]{\smash{{\SetFigFont{9}{10.8}{\familydefault}{\mddefault}{\updefault}{\color[rgb]{0,0,0}$m_2$}%
}}}}
\put(6376,-8194){\makebox(0,0)[b]{\smash{{\SetFigFont{9}{10.8}{\familydefault}{\mddefault}{\updefault}{\color[rgb]{0,0,0}$m_2$}%
}}}}
\put(6901,-6694){\makebox(0,0)[b]{\smash{{\SetFigFont{9}{10.8}{\familydefault}{\mddefault}{\updefault}{\color[rgb]{0,0,0}$m_2$}%
}}}}
\put(6601,-6694){\makebox(0,0)[b]{\smash{{\SetFigFont{9}{10.8}{\familydefault}{\mddefault}{\updefault}{\color[rgb]{0,0,0}$m_1$}%
}}}}
\put(6076,-6394){\makebox(0,0)[b]{\smash{{\SetFigFont{9}{10.8}{\familydefault}{\mddefault}{\updefault}{\color[rgb]{0,0,0}$p_1$}%
}}}}
\put(5701,-6694){\makebox(0,0)[b]{\smash{{\SetFigFont{9}{10.8}{\familydefault}{\mddefault}{\updefault}{\color[rgb]{0,0,0}$p_3$}%
}}}}
\put(5401,-6694){\makebox(0,0)[b]{\smash{{\SetFigFont{9}{10.8}{\familydefault}{\mddefault}{\updefault}{\color[rgb]{0,0,0}$p_1$}%
}}}}
\put(5701,-7294){\makebox(0,0)[b]{\smash{{\SetFigFont{9}{10.8}{\familydefault}{\mddefault}{\updefault}{\color[rgb]{0,0,0}$p_3$}%
}}}}
\put(5701,-7894){\makebox(0,0)[b]{\smash{{\SetFigFont{9}{10.8}{\familydefault}{\mddefault}{\updefault}{\color[rgb]{0,0,0}$p_3$}%
}}}}
\put(5476,-6994){\makebox(0,0)[b]{\smash{{\SetFigFont{9}{10.8}{\familydefault}{\mddefault}{\updefault}{\color[rgb]{0,0,0}$p_3$}%
}}}}
\put(5176,-6994){\makebox(0,0)[b]{\smash{{\SetFigFont{9}{10.8}{\familydefault}{\mddefault}{\updefault}{\color[rgb]{0,0,0}$m_2$}%
}}}}
\put(6376,-7594){\makebox(0,0)[b]{\smash{{\SetFigFont{9}{10.8}{\familydefault}{\mddefault}{\updefault}{\color[rgb]{0,0,0}$p_3$}%
}}}}
\put(6676,-6994){\makebox(0,0)[b]{\smash{{\SetFigFont{9}{10.8}{\familydefault}{\mddefault}{\updefault}{\color[rgb]{0,0,0}$p_3$}%
}}}}
\put(6376,-6994){\makebox(0,0)[b]{\smash{{\SetFigFont{9}{10.8}{\familydefault}{\mddefault}{\updefault}{\color[rgb]{0,0,0}$m_2$}%
}}}}
\put(6076,-6994){\makebox(0,0)[b]{\smash{{\SetFigFont{9}{10.8}{\familydefault}{\mddefault}{\updefault}{\color[rgb]{0,0,0}$m_1$}%
}}}}
\put(6901,-7894){\makebox(0,0)[b]{\smash{{\SetFigFont{9}{10.8}{\familydefault}{\mddefault}{\updefault}{\color[rgb]{0,0,0}$m_3$}%
}}}}
\end{picture}%

%% file: remarks.tex
\section{Concluding Remarks}

We have studied the incorporation of caching in LFTJ by tying an
ordered tree decomposition to the variable ordering. The resulting
scheme retains the inherent advantages of LFTJ (worst case optimality,
low memory footprint), but allows it to accelerate performance based
on whatever memory it decides to (dynamically) allocate. Our
experimental study shows that the result is consistently faster than
LFTJ, by orders of magnitude on large queries, and usually faster than
other state of the art join algorithms.

This work gives rise to several directions for future work. These
include the exploration of caching strategies, finding decompositions
with beneficial caching, extension to general aggregate operators
(e.g., based on the work of Joglekar et
al.~\cite{DBLP:journals/corr/JoglekarPR15} and Khamis et
al.~\cite{DBLP:journals/corr/KhamisNRR15}), utilizing factorized
representations~\cite{DBLP:journals/pvldb/BakibayevKOZ13,DBLP:journals/tods/OlteanuZ15},
and generalizing beyond joins~\cite{DBLP:conf/icdt/Veldhuizen14}.